\documentclass[a4paper,reqno]{amsart}
\usepackage[text={5.2in,8in},centering]{geometry}
\usepackage[backrefs]{amsrefs}
\usepackage{amssymb}

\usepackage{mathrsfs}

\DeclareMathAlphabet{\mathpzc}{OT1}{pzc}{m}{it}

\usepackage{enumerate}
\usepackage{graphicx}
\usepackage{color}
\definecolor{trustcolor}{rgb}{0.71,0.14,0.07}

\numberwithin{equation}{section}
\theoremstyle{plain}
\newtheorem{theorem}{Theorem}[section]

\newtheorem{lemma}{Lemma}[section]

\theoremstyle{remark}

\newtheorem*{quest*}{Question}
\newtheorem*{remark*}{Remark}
\theoremstyle{remark}

\theoremstyle{definition}

\newtheorem*{definition*}{Definition}
\newtheorem*{notation*}{Notation}
\newtheorem*{notations*}{Notations}
\providecommand{\B}{\mathbf}

\providecommand{\C}{\mathcal}

\providecommand{\D}{\mathbb}

\newcommand{\ee}{\mathrm{e}}

\DeclareMathOperator{\dist}{dist}

\DeclareMathOperator{\Const}{Const}

\DeclareMathOperator{\one}{\mathbf{1}}

\DeclareMathOperator{\supp}{{\rm supp}}

\DeclareMathOperator{\diam}{{\rm diam}}

\def\ual{\B{\alpha}}

\def\DN{\D{N}}
\def\DP{\D{P}}
\def\DQ{\D{Q}}
\def\DR{\D{R}}
\def\DT{\D{T}}

\def\DZ{\D{Z}}

\def\cB{\C{B}}
\def\cC{\C{C}}

\def\cF{\C{F}}
\def\cH{\C{H}}

\def\cR{\C{R}}

\def\om{\omega}
\def\Om{\Omega}

\def\Lam{\Lambda}

\def\trho{{\tilde\rho}}

\def\th{\theta}
\def\Th{\Theta}
\def\eps{\epsilon}

\def\ffi{\varphi}

\def\tilTheta{{\widetilde{\Theta}}}

\def\pt{\partial}

\def\lr#1{{\langle #1\rangle}}

\def\tto#1{\smash{\mathop{\,\,\,\, \longrightarrow \,\,\,\, }\limits_{#1}}}

\def\myset#1{{\left\{\,#1\,\right\}}}

\def\thnk{{\th_{n,k}}}

\def\ffink{{\ffi_{n,k}}}

\def\Cnk{{C_{n,k}}}

\def\Thinf{{\Th^{(\infty)}}}

\def\USR{{$\mathbf{(USR)}$}}
\def\DIV{{$\mathbf{(DIV)}$}}

\def\nt{{\widetilde{n}}}

\def\diy{\displaystyle}

\def\ba{\begin{array}{l}}
\def\ea{\end{array}}
\def\bal{\begin{aligned}}
\def\eal{\end{aligned}}
\def\be{\begin{equation}}
\def\ee{\end{equation}}

\def\prth#1{{\mu\left\{\,#1\,\right\}}}

\begin{document}

\title[Grand Ensembles. II. Localization for generic `haarsh' potentials]
{Grand Ensembles of deterministic operators.\\ II. Localization for generic `haarsh' potentials }

\author[V. Chulaevsky]{Victor Chulaevsky}


\address{D\'{e}partement de Math\'{e}matiques\\
Universit\'{e} de Reims, Moulin de la Housse, B.P. 1039\\
51687 Reims Cedex 2, France\\
E-mail: victor.tchoulaevski@univ-reims.fr}

\date{}
\begin{abstract}
We consider a particular class of lattice Schr\"{o}dinger operators with deterministic  potentials depending upon an infinite number of parameters in an auxiliary measurable space. We prove exponential dynamical localization for generic families in the strong disorder regime, using a variant of the Multi-Scale Analysis. In our model,  the potential is generated by a function on a torus which is discontinuous (`haarsh') and constructed with the help of an expansion which reminds Haar's wavelet expansions.

\vskip5mm
\textcolor{blue}
{\textbf{NOTE}: This text is a reduced version of the original manuscript,
originally uploaded in 2009 and revised in 2011.
It is is kept in \textbf{arXiv} to avoid broken references in earlier works.
In a recent preprint
\vskip1mm
"\emph{Uniform Anderson localization, unimodal eigenstates and simple spectra in a class of ``haarsh''
deterministic potentials}"
\vskip1mm
[\texttt{math-ph/1307.7047}], we added new results on the unimodality of the eigenstates,
uniform dynamical localization and simplicity of p.p. spectra.
}
\end{abstract}
\maketitle

\section{Introduction. Formulation of the results.}\label{intro}

      In this paper, we study spectral properties of finite-difference operators, usually called  lattice Schr\"{o}dinger operator (LSO), of the form
\be\label{eq:LSO}
(H f)(x) = \sum_{y:\, \|y-x\|=1}f(y) + gV(x) f(x),\; x,y\in \DZ^d,
\ee
where the function $V:\DZ^d \to \DR$ is usually referred to as the potential; the amplitude $g$ will be assumed positive for the sake of notational brevity. From both physical and purely mathematical point of view, it makes sense to study not an individual operator, but  an entire family of operators
$H(\omega)$ labeled by the points of the phase space of a dynamical system on
some probability space. Moreover, it is convenient to assume the ergodicity of the dynamical
system in question. To define an ergodic family of
operators, one needs:

(i) a probability space $(\Omega, \cF, \DP)$;

(ii) an ergodic dynamical system $T$ with discrete time $\DZ^d$, $d\geq 1$, i.e. a
representation $T: \DZ^d \times \Omega \to \Omega$ of the additive group $\DZ^d$ into the
group of isomorphisms of
$(\Omega, \cF, \DP)$,
$$
T^{x+y} = T^x \circ T^y, \qquad T^x, T^y \in Aut(\Omega, \cF, \DP),
$$
such that any $T$-invariant measurable function on $\Omega$ is a.e. constant;

(iii) a measurable mapping $H$ of the space $\Omega$ into the algebra of
bounded operators acting in the Hilbert space ${\cH} = l^2(\DZ^d)$ verifying for every
$x\in\DZ^d$:
$$
H( T^x(\omega) ) = U^{-x} H(\omega) U^x,
$$
where $(U^x f)(y) = f(y-x)$. The conventional lattice Schr\"{o}dinger operator (LSO) is obtained by setting
$$
H(\om) = \Delta + gV(x;\om),
$$
where $\Delta$ is the nearest-neighbor discrete Laplacian and $V(x;\om)$ is the operator of multiplication by the function
$$
V(x;\om) = v(T^x \om),
$$
with some function $v:\,\Om\to\DR$, which we will call the \textit{hull} of the potential $V$.

An  interesting class of quasi-periodic potentials,  e.g., in one dimension, is obtained when $\Om$ is a torus $\DT^r$ of dimension $r\ge 1$ endowed with the  Haar measure $\DP$ and  the  dynamical system on $\Om$ is  given by
$$
T^x:\, \om  \mapsto  \om + x\alpha \in \DT^r, \quad \alpha\in\DT^r.
$$
As is well-known, this dynamical system is ergodic whenever the frequency vector $\alpha$ has incommensurable (rationally independent) coordinates. Taking a function $v:\DT^r \to \DR$, we can define an ergodic family of quasi-periodic potentials $V:\DZ\to \DR$ by $V(x;\om) := v(T^x \om)$. Multi-dimensional quasi-periodic potentials on $\DZ^n$ can be constructed in a similar way (with the help of $n$ incommensurate frequency vectors $\ual^{j}\in\DR^r, j=1, \ldots, n$).

In this  paper, we do not intend to give an extensive review of prior works on localization properties of quasi-periodic operators. Among the first mathematically rigorous results on the localization phenomenon featured by a one-dimensional discrete Schr\"{o}dinger equation with the single-frequency quasi-periodic potential  of the form $\cos \alpha x$,  $\alpha\in\DR\setminus \DQ$, (also known as Almost Mathieu equation and Harper's equation) we refer to the papers by Sinai \cite{Sin87}  and Fr\"{o}hlich, Spencer and Wittwer \cite{FSW87}. Later, Bourgain, Goldstein and Schlag considered potentials generated by various dynamical systems on a torus $\Om=\DT^\nu$, where the hull $v(\om)$ was assumed analytic; see, e.g., \cite{BG00}, \cite{BGS01}, \cite{BS00}. Recently, Chan \cite{Chan07} proved the  Anderson localization for single-frequency quasi-periodic operators with the hull $v$ of class $C^3(\DT^1)$, using a parameter exclusion technique which is different from presented in this paper.

Below we encapsulate the requirements for the dynamical system in one, mild condition -- that of "uniformly slow" returns of any trajectory $\myset{T^x\om, x\in \DZ^d}$ to small neighborhoods of its starting point $\om\in\Om$; cf. Sect.~\ref{SubSecUSR}.

\subsection{Requirements for the dynamical system}\label{SubSecUSR}

We assume that the underlying dynamical system $T$ on the phase space $\Om$, endowed with a distance $\dist_\Om(\cdot,\cdot)$, satisfies the following condition of \emph{uniformly slow return} (USR, in short):

\USR: $\exists\, A,C\in(0,\infty)\; \;\forall\, \om\in\Om \;\forall\, \lr{x,y}\in\DZ^d$
\be\label{eq:USR}
\begin{array}{lc}
\quad \dist_{\Om}(T^x \om, T^y \om) \ge 4C \|x - y \|^{-A},
\end{array}
\ee

Actually, this condition can be further relaxed so as to admit the lower bound of the form $C e^{-\|x-y\|^\beta}$, with some
$\beta\in(0,1)$ and $C>0$.

In this paper, we consider mainly the case where $\Om = \DT^\nu$, $\nu\ge 1$, and it is technically convenient to define the distance $\dist_\Om[\om', \om''] \equiv \dist_{\DT^\nu}[\om', \om'']$ as follows:
$$
\dist_{\DT^\nu}[(\om'_1, \ldots, \om'_\nu), (\om''_1, \ldots, \om''_\nu)]
= \max_{1 \le i \le \nu} \dist_{\DT^1}[\om'_i, \om''_i],
$$
where $\dist_{\DT^1}$ is the conventional distance on the unit circle $\DT^1$. With this definition, the diameter of a cube of sidelength $r$ in $\DT^\nu$ equals $r$, for any dimension $\nu\ge 1$. The reason for the choice of the phase space $\Om=\DT^\nu$ is that many parametric families of ensembles of potentials $V(x;\om;\th)$ can be made fairly explicit in this case.

We will sometimes work with  the balls
$$
B_r(\om) := \{\om'\in\Om:\, \dist_\Om(\om, \om')\le r\},
$$
which are actually cubes in $\Om$, since $\dist_\om$ is induced by max-norm.

For ergodic rotations of the torus $\DT^\nu$,
$$
T^x \om = \om + x_1 \alpha_1 + \cdots + x_d \alpha_d,
\; x \in\DZ^d, \; \alpha_j\in \DT^\nu, \, 1 \le j \le d,
$$
the USR property reads as a Diophantine condition for the frequency vectors $\ual_j$, which we always assume below.

We will need the following simple consequence of the pointwise separation property \USR\, of the trajectories in the phase space $\Om$.

\begin{lemma}\label{lem:USR.cubes}
Assume the condition \USR\, and let $L>0$ be an integer. Consider a cube $\Lam_L(u)$, $u\in\DZ^d$
then for any $r\le C |\Lam_L(u)|^{-A}$
\be\label{eq:sep.USR.cubes}
\inf_{\om\in\Om} \min_{\lr{x,y}\in\Lam_L(u)} \;
\dist_\Om\left( T^xB_r(\om), T^y B_r(\om)   \right) \ge 2C |\Lam_L(u)|^{-A}.
\ee
\end{lemma}

We also assume a polynomial bound on the rate of divergence of trajectories of the underlying dynamical system.

\DIV: $\exists\, A',C'\in(0,\infty)\; \;\forall\, \om,\om'\in\Om \;\forall\, x\in\DZ^d $
\be\label{eq:DIV}
\begin{array}{lc}
\quad \dist_{\Om}(T^x \om, T^x \om') \le C' \|x \|^{A'} \dist_\Om(\om, \om').
\end{array}
\ee
This condition is obviously fulfilled for the rotations of the torus, as well as for the skew shifts, e.g.,
$$
(\om_1, \om_2) \mapsto (\om_1 + \alpha_1, \om_2 + \om_1 + \alpha_2).
$$

Strongly mixing dynamical systems, like hyperbolic toral automorphisms, require a different approach and have different mechanisms of localization; this subject is beyond the scope of the present manuscript.

\subsection{A general form of Randelette Expansions}

In \cite{C01,C07} we have introduced \textit{parametric families} of ergodic ensembles of operators
$\{H(\om;\th), \om\in\Om\}$ depending upon a parameter $\th\in\Theta$ in an auxiliary space $\Theta$. It is convenient to endow $\Theta$ with the structure of a probability space, $(\Theta, \cB, \DP^{(\th)})$ in such a way that $\th$ be, in fact, an \textit{infinite} family of IID random variables on $\Theta$, providing an infinite number of auxiliary parameters allowing to vary the hull $v(\om;\th)$ locally in the phase space $\Om$. We called such parametric families Grand Ensembles.

In the framework of lattice Schr\"{o}dinger operators, we gave in  \cite{C01,C07} a more specific construction where $H(\om;\th) = H_0 + V(\cdot;\om;\th)$, with $V(x;\om;\th) = V(T^x\om;\th)$ and
\be\label{eq:randel}
v(\om; \th) = \sum_{n=1}^\infty a_n \sum_{k=1}^{K_n} \th_{n,k} \ffi_{n,k}(\om),
\ee
where the family of random variables $\th := (\thnk, n\ge 1, 1 \le k \le K_n)$ on $\Theta$ is IID, and
$\ffink := (\ffink), n\ge 1, 1 \le k \le K_n<\infty)$ are some functions on the phase space $\Om$ of the underlying dynamical system $T^x$. Representations of the form \eqref{eq:randel} were called \textit{randelette}  expansions.

Further, for the purposes of the MSA, it is convenient to assume that
\renewcommand{\labelitemi}{$\bullet$}
\renewcommand{\labelitemii}{$\diamond$}
\begin{itemize}
  \item $\thnk$ have a probability density; e.g. $\thnk$ are uniformly distributed in $[-1,1]$;
  \item the amplitudes $a_n$ of "generations"  $(\thnk, 1\le k \le K_n)$ satisfy
    \begin{itemize}
      \item an upper bound, to ensure  the convergence of the randelette expansion
      \item an appropriate \textit{lower bound}, to ensure that the contribution of the $n$-th generation of
$\thnk$ is sufficient to wriggle the values of the potential $V(T^x \om;\th)$ via the randelettes $\thnk\ffink$  to avoid strong resonances;
    \end{itemize}
  \item $\diam \supp \ffink$ decay rapidly as $n\to \infty$.
\end{itemize}

Putting the amplitude of the $\ffink$ in the coefficient $a_n$, it is natural to assume $|\ffink(\om)|$ to be bounded. Further, in order to control the potential  $V(T^x \om;\th)$ at any lattice site $x\in\DZ^d$ or, equivalently, at every point $\om\in\Om$, it is natural to require that for every $n\ge 1$,  $\Om$ be covered by the union of the sets where at least one function $\ffink$ is nonzero (and, preferably, not too small).

Notice that the dynamics $T^x$ leaves $\th$ invariant.

\subsection{Description of haarsh randelette expansions}

A very particular, yet  interesting case is where randelettes are piecewise constant functions used in the construction of Haar
wavelets\footnote{In fact, the main results of this paper remain true for expansions over the orthogonal Haar wavelets, but we would like to stress that the orthogonality is \textit{not} relevant here.}. For example, if $\Om = \DT^1 = \DR/\DZ$, we set
$$
\ffink(\om) = \one_{\Cnk}(\om), \;\; \Cnk = \left[ 2^{-n}(k-1), 2^{-n}k\right),
\;\; n\ge 1, \; 1 \le k \le K_n = 2^n.
$$
On a torus of higher dimension, one has to replace intervals of length $2^{-n}$ by cubes of side length $2^{-n}$.
Specifically, given an integer $n\ge 1$, for each integer vector $(l_1, \ldots, l_\nu)$ with $1 \le l_j \le 2^n$, consider the cube
$$
\left[ 2^{-n}(l_1-1), 2^{-n} l_1 \right)\times \; \cdots \;
\times \left[ 2^{-n} (l_\nu-1), 2^{-n} l_\nu \right)
\subset \DT^\nu.
$$
These cubes can be numbered, e.g., in the lexicographical order of vectors $(r_1, \ldots, r_\nu)$, and their total number equals
$K_n = 2^{nd}$. We will denote these cubes by $C_{n,k}$, $k=1, \ldots, K_n$.

Next, introduce a countable family of functions on the torus,
$$
\ffink(\om) = \one_{C_{nk}}(\om), \;n\ge 1, \;k=1, \ldots, K_n,
$$
and a countable family of IID random variables $\thnk$ on an auxiliary probability space $\Theta,\cB,\DP^{(\th)}$, uniformly distributed in $[-1,1]$.

Finally, pick a positive number $b \ge 3d$ and set
\be\label{eq:an}
a_n = 2^{-2bn^2}, \; n\ge 1.
\ee
Now define a function $v(\om;\th)$ on $\Om \times \Theta$,
\be\label{eq:randel}
v(\om;\th)  = \sum_{n=1}^\infty a_n \sum_{k=1}^{K_n} \th_{n,k} \ffi_{n,k}(\om),
\ee
which can be viewed as a family of functions $v(\cdot;\th)$ on the torus, parameterized by $\th\in\Theta$, or as a particular case of a "random" series of functions, expanded over a given system of functions $\ffink$ with "random" coefficients.

Observe that the functions $\ffi_{n,k}$ of the same "generation" $n$ and different values of the "cibling index" $k$ are disjoint, in the particular case of "haarsh" randelette expansion, while $|\th_{n,k}|\le 1$, so that
\be\label{eq:cibling.Linf}
\left\| \sum_{k=1}^{K_n} \th_{n,k} \ffi_{n,k}(\cdot) \right\|_{L^\infty(\Om\times\Th}
\le \max_{k} \left\| \ffi_{n,k}(\cdot) \right\|_{L^\infty(\Om)}  = 1.
\ee
As a result, the convergence of the randelette expansion is determined by that of the series
$\sum_n a_n$.

We will call such expansions "haarsh",  making reference to Haar's (Haarsche, in German) wavelets and to the "harsh" nature of the resulting potentials. Constructing a potential "out of flat pieces" is rather unusual in the framework of the localization theory, where, starting from the pioneering mathematical works by Goldsheid, Molchanov and Pastur, all efforts  were usually made  to avoid "flatness" of the potential considered as a function on the phase space of the underlying dynamical system.
Yet, with an infinite number of flat components $\thnk\,\ffink(\om)$, each modulated by its own parameter $\thnk$, we proved in \cite{C01,C07} an analog of Wegner bound for the respective Grand Ensembles $H(\om;\th)$. This was the first indication that such parametric ensembles may feature the phenomenon of Anderson localization.

We use a variant of the Multi-Scale Analysis and study first the spectral properties of finite-volume approximants of the operator $H(\om;\th)$ obtained by its restriction on lattice cubes
$
\Lam_{L_j}(u) = \myset{x\in\DZ^d:\; \|x-u\| \le L_j},
$
with Dirichlet boundary conditions on the "external boundary"
$
\pt^{+} \Lam_{L_j}(u) = \myset{x\in\DZ^d:\; \|x-u\| = L_j + 1}.
$
Here and below, we use the max-norm for vectors $x\in\DR^d$:
$
\diy\|x\|  := \max_{1\le i \le d} |x_i|
$.

The main result of this paper is the following

\begin{theorem}\label{Thm_SL}
Consider a family of lattice Schr\"{o}dinger operators of the form \eqref{eq:LSO} with potential  $V(x;\om;\th) = v(T^x\om;\th)$, where $v(\om;\th)$ is given by the expansion \eqref{eq:randel}, and the dynamical system $T^x$ satisfies conditions \USR\, and \DIV\, for some $A,C<\infty$.

For sufficiently large $g\ge  g_0(C,A)$,  there exists a subset
$\Thinf(g) \subset \Theta$ of measure $\prth{\Thinf(g)} \ge 1 - c(C,A)\, g^{-1}$ with the following property: if $\th\in\Thinf$, then for any $\om\in\Om$
the operator $H(\om;\th)$ has pure point spectrum with exponentially decaying eigenfunctions $\psi_j(\cdot;\om;\th)$:
$$
\forall\, x\in\DZ^d\;\; |\psi_j(x;\om;\th)| \le C_j(\om;\th) e^{-m\|x\|}, \; m = m(g, C, A)>0.
$$
Moreover, there exists $L^*\in\DN$ such that for any bounded measurable function $f$ and
all $x,y\in\DZ^d$ with $\|x-y\|\ge L^*$
$$
 | \langle \delta_x \,|\, f(H(\om,\th)) \,|\,  \delta_y \rangle | \le e^{-m\|x-y\|/2} \, \|f\|_\infty.
$$
\end{theorem}

\section{Partitions and separation bounds for the potential}\label{SecPartitions}

\subsection{Partitions}

For every $n\ge 1$, the supports $C_{n,k} = \supp \ffi_{n,k}$,  $1 \le k \le K_n \}$  generate a partition of the phase space $\Om$:
$$
\cC_n = \myset{ C_{n,k}, 1\le k \le K_n  }.
$$
These partitions form a monotone sequence: $\cC_{n+1}  \prec \cC_n$, i.e., each element of $\cC_n$ is a union of some elements of the partition $\cC_{n+1}$. In the probabilistic language, the (finite) sigma-algebras $\cB_n$ canonically generated by (the elements of) the partitions $\cC_n$ form a monotone family: $\cB_n \subset \cB_{n+1}$.

To each element $C_{n,k}$ of the partition $\cC_n$ corresponds a unique finite sequence of indices
$\kappa(n,k) =(k_1, \ldots, k_n)$ with $k_n=k$ labeling $n$ elements $C_{i,k_i} \supset C_{n,k}$, $1 \le i \le n$, of partitions preceding or equal to $\cC_n$. Further, we associate with the element $C_{n,k}$ a random variable $\xi_{n,k} = \xi_{n,k}(\th)$ relative to the probability space $\Theta$,
$$
\xi_{n,k}(\th) := \sum_{i=1}^n a_i \th_{i, k_i}, \text{ with }  (k_1, \ldots, k_n=k) = \kappa(n,k).
$$
Introduce the approximants of the hull  $v(\om;\th)$ given by \eqref{eq:randel},
\be\label{eq:approx.vN}
v_n(\om;\th) = \sum_{i=1}^n a_i \sum_{k=1}^{K_n} \th_{n,k} \ffi_{n,k}, \; n=1, 2, \ldots \,
\ee
and the corresponding approximants of the potential,
$
V_n(x;\om,\th) = v_n(x;\om,\th).
$
Observe that
\be\label{eq:VN.vN.inf}
\|V - V_n \|_\infty \equiv
\|V - V_n \|_{L^\infty(\DZ^d \times \Om\times\Th)}
\le \|v - v_n \|_{L^\infty( \Om\times\Th)}.
\ee
The random variables $\xi_{n,k}(\th)$ with different $k$ are strongly correlated via the values $\th_{n'}$ with $n'<n$. Nevertheless, the variables  $\th_{n,k}(\th) $, independent for different $k$, bring enough "innovation" and allow to mimick, albeit weakly, various properties of "genuinely random" potentials $V(x;\om)$ with IID values.

In this paper, we consider only functions $\ffi_{n,k}(\om)$ which are indicators of their respective supports, i.e. indicators of the respective partition elements $C_{n,k}$. Therefore, an approximant $v_n(\om;\th)$ can be expressed as follows:
$$
v_n(\om;\th) = \sum_{k=1}^{K_n} \xi_{n,k}(\th) \ffi_{n,k}(\om) = \sum_{k=1}^{K_n} \xi_{n,k}(\th) \one_{C_{n,k} }(\om).
$$
In other words, if $T^x \om \in C_{n,k}$ and $T^y \om \in C_{n,k'}$ with $k\ne k'$, then
\be\label{eq:vn.xi.k.kprime}
 v_n(T^x \om; \th) -  v_n(T^y \om; \th)
= \xi_{n,k}(\th) - \xi_{n,k'}(\th)
\ee
so that
\be\label{eq:V.xi.k.kprime}
|V(x;\om;\th) - V(y;\om;\th)| \ge
 g\,( |\xi_{n,k}(\th) - \xi_{n,k'}(\th) | - \rho_n).
\ee
where
$
\rho_{n} := \|  v - v_n \|_{L^\infty(\Om\times\Th)}.
$

\begin{lemma}\label{lem:approx.Vn}
For any $N\ge 0$,
\be\label{eq:lem.approx.Vn}
\rho_{N}  \le \trho_N := 2^{-2bN} a_N.
\ee
\end{lemma}

\proof

Since $b>1$, for any $N\ge 1$ we have
$$
\bal
\sum_{n=N+1}^\infty a_n & = \sum_{n=N+1}^\infty 2^{-bn^2}
= 2^{-b(2N+1)} 2^{-bN^2} \sum_{i=1}^\infty 2^{-b[(N+i)^2 + b(N+1)^2] }
\\
& \le 2^{-b(2N+1)} a_N \sum_{i=1}^\infty 2^{-i} = 2^{-2bN} a_N.
\eal
$$
Applying \eqref{eq:cibling.Linf}, we conclude that
$$
\|  v - v_N \|_\infty \le \sum_{n=N+1}^\infty a_n \le 2^{-2bN} a_N.
$$
\qedhere

Notice that, for $N$ large, the RHS is \textit{much smaller} than the width $2a_N$ of the distribution of random coefficients $a_N\th_{N,k}$,
$1 \le k \le K_N$ (recall that $\th_{N,k} \sim Unif[-1,1]$). This fact plays an important role in our analysis. Observe also that, since $b > 2d \ge 2$, we have, for all $N\ge 1$,
$$
2^{-bN} a_N \ge 4\, a_N 2^{-2bN}.
$$

\section{ Separation of finite-volume spectra}
\label{sec:spec.spectra}

\subsection{Initial scale bounds}
$\,$\par\vskip1mm\textcolor{blue} {Subsection removed. See more complete results in math-ph/1307.7047.}

\subsection{An arbitrary scale}
$\,$\par\vskip1mm\textcolor{blue} {Subsection removed. See more complete results in math-ph/1307.7047.}

\subsection{A deterministic Minami-type estimate}

For each given integer $L\ge 1$ we will consider again two partitions of the torus used in the proof of Lemma \ref{lem:cor.lem.DLom}:
\begin{itemize}
  \item  into cubes $Q'_j(r)$, $r = [L^{-A-A'}/2]$ with centers $\om_i$ of the form
$$
\om_i = \left[ l_1 r^{-1}, \ldots, l_\nu r^{-1} \right),
\; l_1, \ldots, l_\nu\in [[0, (2r)^{-1}-1]];
$$

  \item into cubes $Q_j(R)$ of radius $3R$, with $R = [L^{-A'}/6]$, each partitioned into $3^\nu$ adjacent cubes of radius $R$.
\end{itemize}
The notations $Q_j(R)$  and $Q'_j(r)$ will be used throughout this section.
\smallskip

Recall that, for any $x\in\Lam_L(0)$, if $T^x \om_i\in Q_{i,1}(R)$ and
$\om\in Q'_i(r)$, then $T^x\om \in Q_i(R)$.

\begin{lemma}\label{lem:Minami.1cube}
Let $A,C>0$ be the constants from Eqn \eqref{eq:USR} and $A'>0$ be the constant from Eqn \eqref{eq:DIV}. Fix a cube $Q'_k(r)\subset\DT^\nu$, a lattice cube $\Lam_L(u)$, and consider the operators $H^{(L)}(\om;\th) = H_{\Lam_L(u)}(\om;\th)$ with $\om\in Q'_k(r)$ and their eigenvalues
$E_i = E_i(\om;\th)$.  Then for any $s\ge 2g\cdot 2^{-b\nt(L)}a_{\nt(L)}$
\be
\prth{\th:\,\exists\, \om\in Q'_k(r)\;\; \min_{\lr{i,j}} |E_i - E_j|\le s  }
\le \Const \, g^{-1} L^{2d} e^{2A \ln^2L} s.
\ee
\end{lemma}

For any integer $L\ge L_0$ and any $s>0$, introduce the subset of the space $\Th$,
\be
\cR(L,s) =
\left\{ \th:\,\exists\,\om\;\; \min_{\lr{i,j} } |E_i(\om;\th) - E_j(\om;\th)|\le s \right\}.
\ee

\begin{lemma}\label{lem:Minami}
Under the assumptions of Lemma \ref{lem:Minami.1cube},
\be
\prth{\cR(L,s) } \le Const(C,A)\, g^{-1}  L^{A+A'+2d} \, e^{ 2A \ln^2 L}  \, s.
\ee
If, in addition, $L\ge L_0 \ge e^{(A+A'+2d)/A}$, then
\be
\prth{\cR(L,s) } \le Const(C,A)\,   g^{-1} \, e^{ 3A \ln^2 L}  \, s.
\ee
\end{lemma}

\subsection{Recursive construction of "good" parameter subsets $\Th^{(j)}$}

Introduce the following  events relative to the probability space $\Theta$:
$$
\tilTheta^{(j)} = \{\theta:\, D(L_{j-1}, \theta)  \ge 4\delta_{j} \} \setminus \cR(L_{j}, \eps_j),
\quad j\ge 0,
$$
and, recursively,
$$
\Theta^{(j)} = \tilTheta^{(j)} \cap \Theta^{(j-1)},\quad
\Thinf = \bigcap_{j=0}^\infty \Theta^{(j)}.
$$

Now we can formulate to the key result on spectral spacings in finite volumes.

\begin{lemma}\label{Lem_SepSpecAnyScale}
Let $j\ge 1$ and consider the scales $L_j$, $L_{j+1}$ defined in Eqn (3.2).
Then

\be\label{eq:bound.Th.j}
\prth{\Theta^{(j)} \setminus \Theta^{(j+1)} }
\le \Const \,    e^{ - \ln^2 L_j }
\ee
and, therefore,
\be
\prth{\Theta^{(\infty)} } \tto{L_0\to\infty} 1.
\ee
\end{lemma}

\section{Decay of Green functions in finite boxes}\label{SecDecayGF}

$\,$\par\vskip1mm
\textcolor{blue}{
[This section is removed. See more complete results in math-ph/1307.7047.
}

\section*{Acknowledgements}

I thank Misha Goldstein for fruitful discussions of localization techniques for deterministic operators and his hospitality during my visit to the University of Toronto in Spring 2009. I also thank Gian Michele Graf, G\"{u}nter Stolz and Jean Bellissard for stimulating discussions of Wegner--Minami type bounds.


\end{document}